\documentclass[aps,prb,twocolumn,showpacs,superscriptaddress]{revtex4}
\usepackage{graphicx}
\begin{document}

\title{Interfacial thermal transport in atomic junctions}
% author.
     \author{Lifa~Zhang}
     \affiliation{Department of Physics and Centre for Computational Science and Engineering,
     National University of Singapore, Singapore 117542, Republic of Singapore }
     \author{Pawel~Keblinski}
      \affiliation{Department of Materials Science and Engineering, Rensselaer Polytechnic Institute, New York, 12180, USA. }
     \author{Jian-Sheng~Wang}
     \affiliation{Department of Physics and Centre for Computational Science and Engineering,
     National University of Singapore, Singapore 117542, Republic of Singapore }
     \author{Baowen~Li}
    \altaffiliation{Electronic address: phylibw@nus.edu.sg}
    \affiliation{NUS Graduate School for Integrative Sciences and Engineering,
      Singapore 117456, Republic of Singapore}
       \affiliation{Department of Physics and Centre for Computational Science and Engineering,
     National University of Singapore, Singapore 117542, Republic of Singapore }
\date{30 Oct 2010, Revised 10 Jan 2011 }
\begin{abstract}
{We study ballistic interfacial thermal transport across atomic
junctions. Exact expressions for phonon transmission coefficients
are derived for thermal transport in one-junction and two-junction
chains, and verified by numerical calculation based on a
nonequilibrium Green's function method. For a single-junction case,
we find that the phonon transmission coefficient typically decreases
monotonically with increasing freqency. However, in the range between equal frequency spectrum and equal acoustic impedance, it increases first then decreases, which explains why the Kapitza resistance calculated from the acoustic mismatch model is far larger than the experimental values at low temperatures.
The junction thermal conductance reaches a maximum when the
interfacial coupling equals the harmonic average of the spring
constants of the two semi-infinite chains. For three-dimensional
junctions, in the weak coupling limit, we find that the conductance
is proportional to the square of the interfacial coupling, while for
intermediate coupling strength the conductance is approximately
proportional to the interfacial coupling strength.  For two-junction
chains, the transmission coefficient oscillates with the frequency
due to interference effects. The oscillations between the two
envelop lines can be understood analytically, thus providing
guidelines in designing phonon frequency filters.}
\end{abstract}

\maketitle

\section{Introduction}

In the past decade there has been a significant research focus on
thermal transport in micro scale\cite{Dhar}. Several conceptual
thermal devices, such as thermal rectifiers/diodes, thermal
transistors, thermal logical gates, and thermal memory
\cite{rectifiers,transistor,logicgate,memory}, have been proposed,
which, in principle, make it possible to control heat due to phonons
and process information with phonons.  The issue of quantum thermal
transport in nanostructures was also addressed \cite{wangjs2008}. In
this context, the critical information is in phonon transmission
coefficients that in quasi-one-dimensional atomic models can be
calculated by transfer matrix method
\cite{tong1999,macia2000,cao2005,antonyuk2005}. However, the
evaluation of the transfer matrix may be numerically unstable,
particularly when the system size becomes large. Alternatively,
nonequilibrium Green's function (NEGF) method is an efficient way to
calculate the transmission coefficient\cite{negfref}. Unfortunately,
both of these two methods are numerical in nature and do not give
analytical expressions.

For thermal transport and control, the interfacial thermal scattering process is becoming increasingly important, especially in practical devices. Two theories, acoustic mismatch model \cite{little1959} and the diffuse mismatch model \cite{swartz1989}, have been proposed to study the mechanism of the thermal interfacial resistance. However, both models offer limited accuracy in nanoscale interfacial resistance predictions \cite{stevens2005} because they neglect atomic details of actual interfaces.  A scattering boundary method within the lattice dynamic approach was first proposed by Lumpkin and Saslow to study the Kapitza conductance in a one-dimensional (1D) lattice \cite{lumpkin1978}, and was then applied to calculate the Kapitza resistance in two- and three-dimensional (3D) lattices \cite{paranjape1987,young1989}. This method can predict thermal interfacial conductance between heterogeneous materials with full consideration of the atomic structures in the interface.  Recently, this method was applied to study the ballistic thermal transport in
nanotube junctions\cite{wang2006},  spin chains\cite{zhang2008}, and honeycomb lattice ribbons \cite{cuansing2009}.

In this paper we give an explicit analytical expression of transmission coefficient obtained through the scattering boundary
method, and use it to study the interfacial thermal transport across atomic junctions.  First, in Sec.~\ref{secmod}, we introduce a model in which two semi-infinite 1D atomic chains are coupled either via a point junction or an extended junction region. By using the boundary scattering method we derive the exact expressions for phonon transmission coefficients for thermal transport
in one-junction and two-junction chains in Sec.~\ref{secanal}. The role of various parameters on the junction conductance is analyzed and discussed in Sec.~\ref{secrd}. In section ~\ref{secrd} we also estimate the interfacial conductance between two 3D solids. In Sec.~\ref{secnegf}, we introduce briefly the NEGF method, and use it to verify the results from analytical formulae for the thermal transport in our model.  A short summary is presented in Sec.~\ref{seccond}.

\begin{figure}[h]
\includegraphics[width=3.4 in,angle=0]{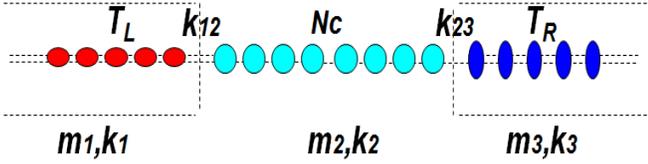}%
 \caption{ \label{fig1model}(color online) A schematic
representation of the 1D atomic chain model. The size of the center part is $N_C=8$. The left and right regions are two semi-infinite harmonic atomic chains at different temperatures $T_L$ and $ T_R$. The three parts are coupled by harmonic springs with constant strength $k_{12}$ and $k_{23}$; all of which are harmonic chains with mass  and spring constant as $m_1,k_1$, $m_2,k_2$ and $m_3,k_3$, respectively. }
\end{figure}

\section{Model}
\label{secmod} The one-dimensional atomic chain consists of three
parts: two semi-infinite leads and  an center region (see
Fig.~\ref{fig1model}). The two leads are in equilibrium at different
temperatures $T_L $ and $T_R$. The three parts are coupled  by
harmonic springs with constant strength $k_{12}$ and $k_{23}$; all
of which are harmonic chains with mass and spring constants
$m_1,k_1$, $m_2,k_2$ and $m_3,k_3$, respectively. So the total
Hamiltonian can be written as
\begin{equation}\label{eqham}
H = \sum\limits_{\alpha  = 1,2,3} {H_\alpha  }  + \frac{1}
{2}k_{12} (x_{1,1}  - x_{2,1} )^2  + \frac{1}
{2}k_{23} (x_{2,N_c }  - x_{3,1} )^2;
\end{equation}
here,
\begin{equation}
 H_\alpha   = \sum\limits_{i=1}^{N_\alpha} {\frac{1}
{2}m_\alpha  \dot x_{\alpha ,i}^2  + \sum\limits_{i=1}^{N_\alpha-1}\frac{1}
{2}k_\alpha  (x_{\alpha ,i}  - x_{\alpha ,i + 1} )^2 }.
\end{equation}
Where $x_{\alpha ,i}$ is the relative displacement of i-th atom in
$\alpha$-th part.  If there is no center part, that is, the two
semi-infinite leads connected directly by $k_{12}$,  then by setting
$\alpha = 1,2$ and $k_{23}=0$ in Eq.~(\ref{eqham}), we can obtain
the corresponding Hamiltonian. For the semi-infinite leads, $N_\alpha=\infty$.

\section{Analytical Solution from the Scattering Boundary Method}
\label{secanal}
Heat current flowing from left to right
through a junction connecting two leads kept at different
equilibrium heat-bath temperatures $T_L$ and $T_R$ is given
by the Landauer formula \cite{wangjs2008}
\begin{equation}\label{eqflux}
I = \frac{1}{{2\pi }}\int_{0}^\infty
{\hbar\omega \;\bigl[f_L (\omega ) - f_R (\omega )\bigr] T[\omega
]} d\omega,
\end{equation}
which allows us to develop the junction conductance formula
\begin{equation}\label{eqcond}
\sigma = \frac{1}{{2\pi }}\int_{0}^\infty {d\omega\;\hbar\omega\,
 T[\omega]\frac{\partial f(\omega)}{\partial T}},
\end{equation}
here, $ f_{L,R}  = \{ \exp [\hbar \omega /(k_B T_{L,R} )] - 1\} ^{ - 1}
$ is the Bose-Einstein distribution for phonons, and $T[\omega]$ is the frequency dependent transmission coefficient.
Therefore, the key step for the thermal transport characterization is to calculate the transmission coefficients.

We first consider a point-junction case, that is, two semi-infinite
harmonic chains connected by a spring with constant strength
$k_{12}$. We assume a wave solution transmitting from the left lead
to the right lead.  We label the atoms as $-\infty ,\cdots ,-
1,0,1,2, \cdots , + \infty$. Atoms $0$ and $1$ are connected by
$k_{12}$ spring. An incident wave from left is assumed as
$x_I=\lambda _1^j e^{ - i\omega t} $. When it arrives at the
interface, it will be partially reflected and partially transmitted.
The reflected wave amplitude is $x_R  = r_{12} \lambda _1^{ - j} e^{
- i\omega t} $ and the transmission wave can be written as $x_T  =
t_{12} \lambda _2^{j - 1} e^{ - i\omega t}$. So at each atom we have
$\cdots ,\;x_{ - 1}  = (\lambda _1^{ - 1}  + r_{12} \lambda _1 )e^{
- i\omega t} ,\;x_0  = (1 + r_{12} )e^{ - i\omega t}$,  $
  x_1  = t_{12} e^{ - i\omega t} ,\;x_2  = t_{12} \lambda _2 e^{ - i\omega t} ,\; \cdots $. Here, $\lambda _j=e^{iq_ja_j}$, $q_j$ is the wave vector,
  $a_j$ is the interatomic spacing.  For the atom in the $j-th$ part, we can have the equation of motion as
\begin{equation}
m_j \frac{d^2 x_{j,n}}{dt^2}=k_j(x_{j,n+1}-x_{j,n})+k_j(x_{j,n}-x_{j,n-1}),
\end{equation}
each wave transport separately and satisfies such equation. Thus
 $\lambda _j$  satisfies the dispersion relation of the corresponding lead as
\begin{equation}\label{eqdspr}
\omega ^2 m_j  =  - k_j \lambda _j^{ - 1}  + 2k_j  - k_j \lambda _j.
\end{equation}
The quadratic equation has two roots. Which one should we choose?
Replacing $\omega$ with $\omega + i\eta $, $\eta  = 0^ +$, none of
the eigenvalues $\lambda$ will have modulus exactly 1. We find for
the traveling waves \cite{velev2004}
\begin{equation}
|\lambda|=1-\eta \frac{a}{v},
\end{equation}
thus the forward moving waves with group velocity $ v > 0$
have $|\lambda| < 1$. Therefore we should take the one with $|\lambda|<1$ of the two roots which are given as
\begin{equation}\label{eqlmbd}
\lambda _j  = \frac{{ - h_j  \pm \sqrt {h_j ^2  - 4} }}
{2},\;\;\;h_j  = \frac{{m_j }}
{{k_j }}(\omega  + i\eta )^2  - 2.
\end{equation}

From the scattering boundary method, the coefficients $r_{12}$, $t_{12}$  can be obtained from the continuity condition at the interface as:
\begin{eqnarray}
  \omega ^2 m_1 x_0  =  - k_1 x_{ - 1}  + (k_1  + k_{12} )x_0  - k_{12} x_1;  \\
  \omega ^2 m_2 x_1  =  - k_{12} x_0  + (k_{12}  + k_2 )x_1  - k_{12} x_2 .
 \end{eqnarray}
Finally we can get the transmission coefficient as
\begin{equation}\label{eqtran1}
 T[\omega ] = 1 - |r_{12} |^2  = 1 - |r_{21} |^2,
\end{equation}
here,
\begin{equation}\label{eqrflt}
r_{ij}  = \frac{{k_i (\lambda _i  - 1/\lambda _i )(k_j  - k_{ij}  - k_j /\lambda _j )}}
{{(k_i  - k_{ij}  - k_i /\lambda _i )(k_j  - k_{ij}  - k_j /\lambda _j ) - k_{ij}^2 }} - 1.
\end{equation}
Of course, we can also use $t_{12}$ to express $T[\omega]$ as $\frac{{m_2 v_2 /a_2 }}
{{m_1 v_1 /a_1 }}|t_{12} |^2 $, here the group velocity $v_i  = \frac{{d\omega }}
{{dq_i }} = \frac{{a_i }}{2}\sqrt {\frac{{4k_i }}{{m_i }} - \omega ^2 }$, which is derived from the dispersion relation given by Eq.~(\ref{eqdspr}). Thus, the transmission coefficient can also be expressed as
\begin{equation}\label{eqtran1t}
T[\omega ] = \frac{{\sqrt {4k_2 m_2  - \omega ^2 m_2^2 } }} {{\sqrt {4k_1 m_1  - \omega ^2 m_1^2 } }}|t_{12} |^2,
\end{equation}
here
\begin{equation}\label{eqtrsm}
t_{ij}  = \frac{{ - k_{ij} k_i (\lambda _i  - 1/\lambda _i )}}
{{(k_i  - k_{ij}  - k_i /\lambda _i )(k_j  - k_{ij}  - k_j /\lambda _j ) - k_{ij}^2 }}.
\end{equation}

For the long-wave limit, that is, $\omega = 0^+$, we get $r_{ij}  = \frac{{\sqrt {k_i m_i }  - \sqrt {k_j m_j } }}
{{\sqrt {k_i m_i }  + \sqrt {k_j m_j } }}$; and the transmission is
\begin{equation}\label{eqtran0}
T[0^ +] = \frac{{4\sqrt {k_1 m_1 k_2 m_2 } }}
{{(\sqrt {k_1 m_1 }  + \sqrt {k_2 m_2 } )^2 }}.
\end{equation}
This result is consistent with the one obtained for the acoustic mismatch model, i.e.,
 $T= \frac{4Z_1 Z_2 }
{(Z_1+ Z_2)^2}.$ \cite{little1959} Where the acoustic impedance is $Z_i=\rho_i v_i=(m_i/a_i) v_i$, and $Z_i(\omega=0^+)=\sqrt {k_i m_i}$. We note that in acoustic mismatch model the transmission coefficient is frequency independent, and in reality it only applies in  the limit of low frequency/long wavelengths. In this case the phonon sees the interface only as a discontinuity between two semi-infinite media and the transmission does not depend on the coupling spring strength $k_{ij}$. If the two leads have the same acoustic impedance for long wave limit, then $T[0^ +] =1$; otherwise $T[0^ +]<1$.

For a two-junction case, which is shown in Fig.~\ref{fig1model}, the transmission wave will be reflected and transmitted by the second boundary, leading to multiple reflections. Finally the total transmitted wave function is obtained as a superposition of multiple reflections and transmissions, resulting in the transmission coefficient through the center part
\begin{equation}\label{eqtran2}
T[\omega] = \frac{{(1 - |r_{12} |^2 )(1 - |r_{23} |^2 )}}
{{|1 - r_{23} r_{21} \lambda _2^{2(N_C - 1)} |^2 }},
\end{equation}
here $r_{ij}$ and $\lambda _i$ are determined by Eq.~(\ref{eqrflt}) and Eq.~(\ref{eqlmbd}); $N_C$ is the number of atoms in the center atomic chain. From this expression, we can find that the transmission coefficient oscillates with frequency, and is between the envelope lines of maximum and minimum transmission, which are $
    T_{{\rm max} } [\omega ] = (1 - |r_{12} |^2 )(1 - |r_{23} |^2 )/
(1 - |r_{23} r_{21} |)^2$
for constructive interference and $
 T_{{\rm min} } [\omega ] = /(1 - |r_{12} |^2 )(1 - |r_{23} |^2 )/
(1 + |r_{23} r_{21} |)^2$  for destructive interference.
\begin{figure}[t]
\includegraphics[width=3.0 in,angle=0]{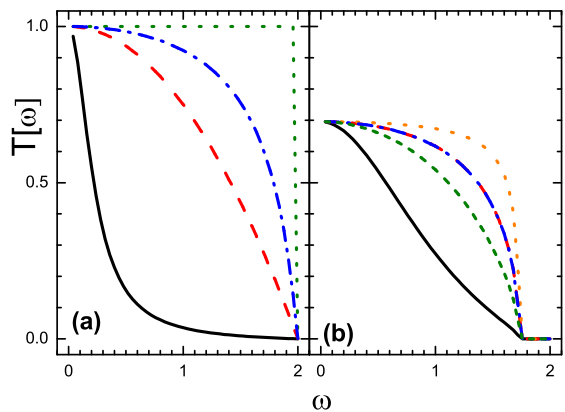}%
 \caption{\label{fig2} (color online) The transmission coefficient vs frequency $\omega$ for different interface coupling $k_{12}$ in one-junction chains. (a) shows the transmission in one junction connected by the same semi-infinite atomic chains with $k_1=k_2=1.0,\; m_1=m_2=1.0$; the solid, dashed, dotted and dash-dotted lines correspond to $k_{12}=0.1$, 0.5, 1.0 and 2.0, respectively. (b) shows the transmission in one junction connected by two different semi-infinite atomic chains with $k_1=1.0,\; m_1=1.0,\; k_2=3.0$ and $m_2=4.0$; the solid, dashed, dotted, dash-dotted and shot-dashed lines correspond to $k_{12}=0.5$, 1.0, 1.5, 3.0 and 8.0, respectively.}
\end{figure}

\begin{figure}[b]
\includegraphics[width=3.0 in,angle=0]{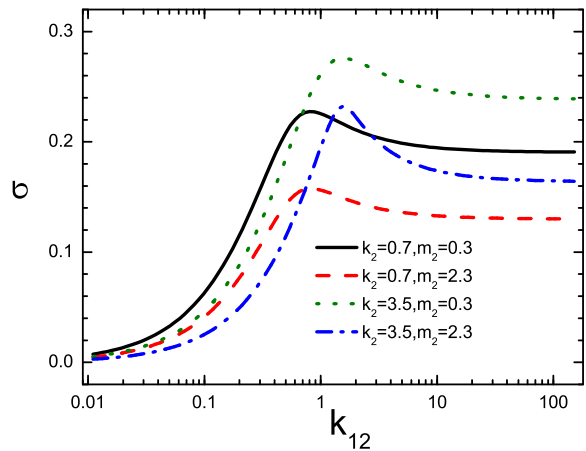}%
 \caption{\label{fig3}(color online) The thermal conductance vs interface coupling $k_{12}$ in point-junction model. Here, $k_1=1.0,\; m_1=1.0$.}
\end{figure}

\section{Results and Discussions}
\label{secrd}

\subsection{Thermal transport in 1D one-junction chains }

In Sec.~\ref{secanal}, we have derived the analytical expressions for the phonon transmission coefficient for point-junction and extended-junction (two point junction) cases Eq.~(\ref{eqtran1}), Eq.~(\ref{eqrflt}) and Eq.~(\ref{eqtran2}) by using the scattering boundary method. Using these analytical expressions, we analyze the role of various parameters on the thermal transport in one- and two- point junctions.

Figure~\ref{fig2} shows the transmission coefficient as a function of frequency for a different interface spring constant $k_{12}$ for the point-junction model. The maximum frequency at which the transmission coefficient is above zero is equal to the minimum of $2\sqrt{k_1/m_1}$ and $2\sqrt{k_2/m_2}$. In Fig.~\ref{fig2}(a), the two semi-infinite atomic chains have the same mass and spring constant. When the interface coupling $k_{12}$ equals to that of the chains, the transmission is equal to one in the whole frequency domain, because of the homogeneity of the chain structure. If $k_{12}$ increases or decreases, the transmission coefficient decreases. If we set $k_1/m_1=k_2/m_2$, the transmission coefficient exhibits similar behavior, the only difference is that the transmission coefficient changes to the value obtained by Eq.~(\ref{eqtran0}).
In Fig.~\ref{fig2}(b), the two semi-infinite atomic chains have different masses and spring constants. The transmission decreases with increased frequency for all the coupling values $k_{12}$. Also, it appears that for a given frequency the transmission is maximized for a $k_{12}$ value residing between $k_1$ and $k_2$.   From Eq.~(\ref{eqtran1}) and Eq.~(\ref{eqrflt}), $T[\omega]=0$, if $k_{12}=0$; and $T[\omega]$ has definite value $1 - |\frac{{k_1 (\lambda _1  - 1) - k_2 (1 - \lambda _2^{ - 1} )}}{{k_1 (1 - \lambda _1^{ - 1} ) + k_2 (1 - \lambda _2^{ - 1} )}}|^2$, if $k_{12}=\infty$.
\begin{figure}[t]
\includegraphics[width=3.0 in,angle=0]{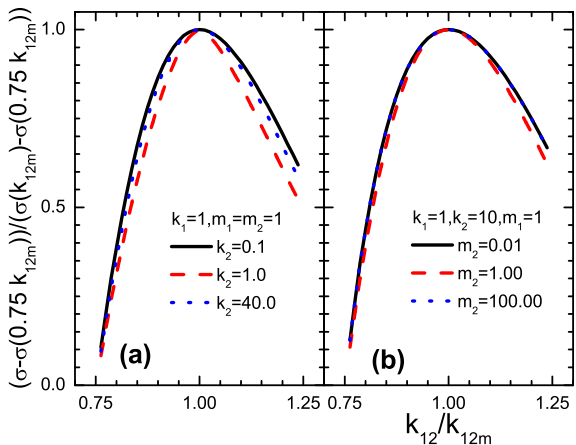}%
 \caption{ \label{fig4}(color online)  The thermal
conductance vs the ratio of $k_{12}/k_{12m}$ in one-junction atomic
chain. Here $k_{12m}$ is the harmonic average of the spring
constants of the two semi-infinite leads. (a) $k_1=1.0$,
$m_1=m_2=1.0$; the solid, dashed, and dotted lines correspond to $k_{2}=0.1$, 1.0, and 40.0, respectively. (b) $k_1=1.0$, $m_1=1.0$, $k_2=10.0$; the solid,
dashed, and dotted lines correspond to
$m_{2}=0.01$, 1.0, and 100.0, respectively. }
\end{figure}

The maximum transmission concept results in the maximum junction conductance as shown in Fig.~\ref{fig3}. With the increasing of $k_{12}$, we find that the conductance will first increase, then arrive at maximum value, and then slightly decrease and at last it will tend to a constant. We find that the maximum transmission or conductance occurs at $k_{12}$ given by
\begin{equation}\label{eqk12m}
 k_{12}=k_{12m}=\frac{2k_1 k_2}{k_1+k_2},
\end{equation}
i.e., when the coupling spring stiffness is equal to the harmonic average of spring connecting atoms in the two semi-infinite chains. In Fig.~\ref{fig4}, we show the thermal conductance vs the ratio of $k_{12}$ and $k_{12m}$.  For the two semi-infinite chains with the same mass $m_1=m_2$, the maximum conductance occurs exactly at $k_{12m}$. If the two leads have different masses $m_1\neq m_2$, the maximum conductance is almost exactly at the $k_{12m}$ point, for mass ratios ranging from 0.01 to 100.
\begin{figure}[t]
\includegraphics[width=3.0 in,angle=0]{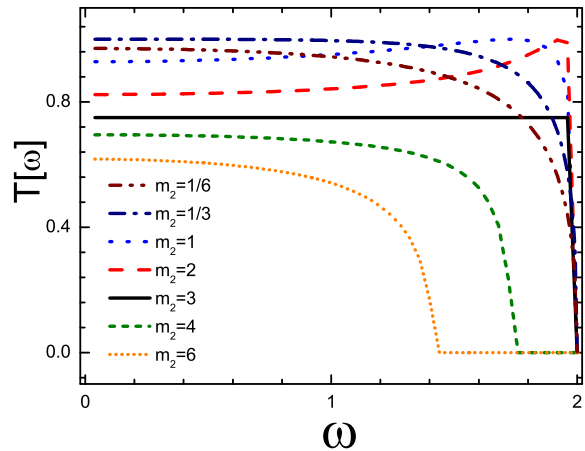}%
\caption{\label{fig5}  (color online)  The transmission
coefficient vs frequency for different mass ratios $m_2/m_1$ at the
interface coupling $k_{12m}$. Here, $k_1=1.0$,
$k_2=3.0$, $k_{12}=k_{12m}=1.5$ and $m_1=1.0$.   }
\end{figure}
\begin{figure}[b]
\includegraphics[width=0.9\columnwidth,angle=0]{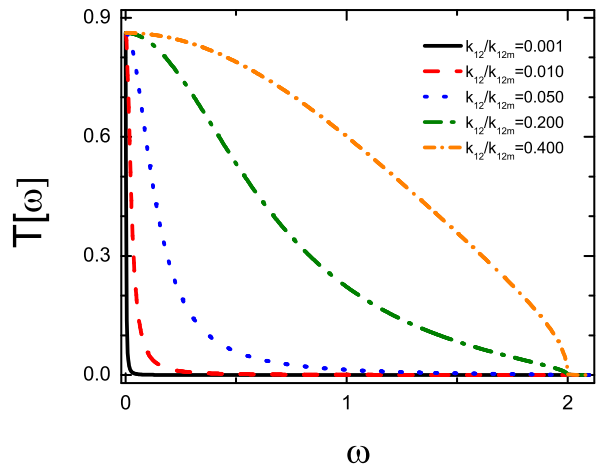}%
 \caption{\label{fig6}(color online) The transmission
coefficient vs frequency for different interface coupling $k_{12m}$.
Here, $k_1=1.0, m_1=1.0$, $k_2=0.7,m_2=0.3$.  }
\end{figure}

In Fig.~\ref{fig5}, we show the curves of the transmission  as a
function of frequency for interface coupling equal to $k_{12m}$. If
$k_1/m_1=k_2/m_2$, that is, when both chains have the same frequency
spectrum of $[0, 2\sqrt{k_1/m_1}]$, the transmission equals to a
constant $T[\omega]=T[0^ +]$,  which can be seen from the solid line
in Fig.~\ref{fig5}, and which is consistent with Fig.~\ref{fig2}(a).
Thus for chains with matched spectra the transmission is frequency
independent. Let us now fix $k_1, k_2$ and $k_2$, and decrease
$m_2$. In the range between the point of equal-spectrum ($\omega_m=k_1/m_1=k_2/m_2$) and the one of equal-impedance ($Z(\omega=0^+)=k_1m_1=k_2m_2$), the transmission will
first increase with frequency and then decrease. Otherwise, there is
a monotonic decrease. The former behavior is quite interesting, as
one expects that the transmission should be the largest in the long
wavelength limit. For highly dissimilar materials, the transmission coefficient in the whole frequency range is much larger than that in the long wave limit $T[\omega=0^+]=\frac{4Z_1 Z_2 } {(Z_1+ Z_2)^2}$,  thus the real conductance is far larger than that calculated from the acoustic mismatch model.  This result explain why the interfacial resistance calculated from the acoustic mismatch model is far lager than the experimental value measured at low temperatures, where the phonon transport can be regarded as ballistic transport.

\begin{figure}
\includegraphics[width=3.0 in,angle=0]{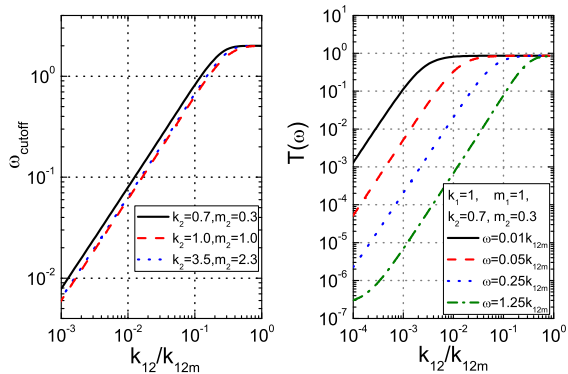}%
 \caption{ \label{fig7}(color online)(a) The cutoff
frequency vs interface coupling for 1D one-junction atomic chains.
The parameters are: $k_1=1.0$, $m_1=1.0$. (b) The transmission as
function of interface coupling for 1D one-junction atomic chains.
The parameters are: $k_1=1.0$, $m_1=1.0$,  $k_2=0.7$, $m_2=0.3$   }
\end{figure}
In many real interfaces, interface coupling is very weak, that is,
the $k_{12}$ is less than $k_{12m}$. So it is desirable to study the
thermal transport in atomic chains in the weak coupling limit.
Figure \ref{fig6} shows the transmission coefficient as function of
interface coupling. In the weak coupling limit, with the frequency
increasing, the transmission decreases rapidly to zero, so the
frequency region where phonons are effectively transmitted is very
narrow.  With interface strength increasing, more and more modes
contribute to the transmission and the phonon transmission window
widens. If the interface coupling increases further, that is
$k_{12}/k_{12m}>0.1 $, out of the weak interface coupling limit, all
the phonons contribute to the transmission. The only further change
with increasing $k_{12} $ is the actual values of the transmission
coefficients increase. In Fig.~\ref{fig7}(a), we show the
transmission cutoff frequency as function of the interface coupling.
Here, we define the cutoff frequency $\omega_{\rm cutoff}$ at which
the transmission $T(\omega_{\rm cutoff})=0.1 T(0^+)$. We find that
the cutoff frequency shows linear dependance on interface coupling
in the weak coupling limit $k_{12}<0.1 k_{12m}$. If the interface
strength increase further, the cutoff frequency is saturated. In
Fig.~\ref{fig7}(b), we show the transmission as function of
interface coupling for several different phonons. We find that in
the weak interface coupling region, the transmission is proportional
to the square of the interface coupling, which is consistent with
the formulas Eq.~(\ref{eqtran1t}) and Eq.~(\ref{eqtrsm}).
\begin{figure}[t]
\includegraphics[width=3.0 in,angle=0]{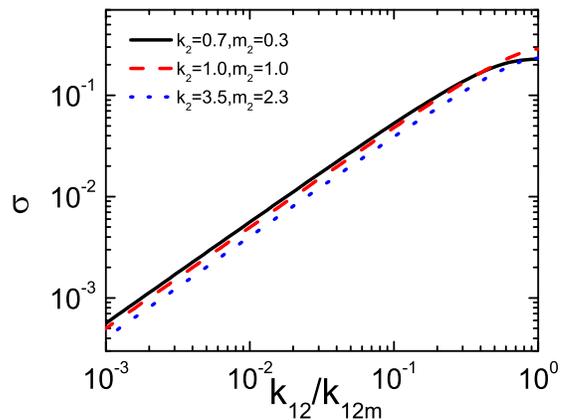}%
 \caption{\label{fig8}(color online) The thermal
conductance vs interface coupling for 1D point-junction atomic
chains. The parameters are: $k_1=1.0$, $m_1=1.0$.  }
\end{figure}
In the weak interface coupling region, for the 1D atomic
one-junction chains, it is shown that the thermal conductance is
linear with the interface coupling (see Fig.~\ref{fig8}). If we
strengthen the interface coupling between the two chains, the
conductance will be linearly enhanced. For different mismatched
chains, the absolute values of the conductance are different, but
dependence on the coupling strength is the same.
\begin{figure}[t]
\includegraphics[width=3.0 in,angle=0]{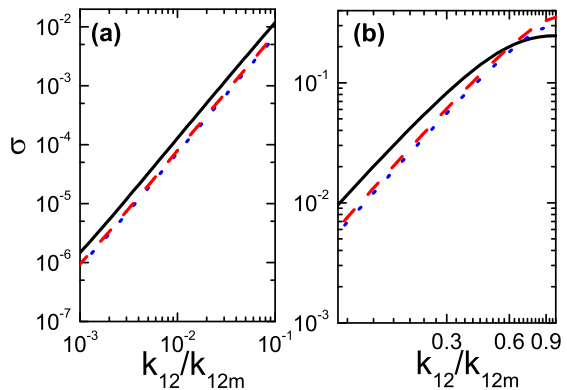}%
\caption{ \label{fig9} (color online) The thermal
conductance  vs interface coupling for 3D one-junction atomic
chains. The parameters are the same with Fig.~\ref{fig8}. (a)
Interface coupling is far less than the coupling $k_{12m}$:
$k_{12m}/k_{12}=0.001-0.1$; (b) Interface coupling is in the region
of $0.1k_{12m}\sim 0.9k_{12m}$. }
\end{figure}

\subsection{Thermal transport in 3D single-interface structures }

The thermal conductance Eq.~(\ref{eqcond}) can also be written as \cite{hopkins2009}:
\begin{equation}\label{3D}
 \sigma = \int_{0}^\infty {d\omega\;\hbar\omega\,
 T[\omega]\frac{\partial f(\omega)}{\partial T}v(\omega) D(\omega)},
\end{equation}

because of $ v(\omega)=\partial \omega / \partial k$  and phonon
density of states in 1D structure, $D(\omega)=1/(2\pi v)$, we can
obtain Eq.~(\ref{eqcond}). In order to estimate the behavior of the
interfacial thermal transport across interfaces in 3D structures, we
only need to change the phonon density of states in the above
equation.  Because the density of states for 3D structure within the
Debye approximation is $D(\omega)\sim \omega^2$, therefore we can
replace $\omega$ with $\omega^3$ in Eq.~(\ref{eqcond}); the thermal
conductance as a function of the coupling strength is shown in
Fig.~\ref{fig9}. From Fig.~\ref{fig9}(a), we find that in the weak
interface limit, conductance is proportional to the square of
interface coupling, which is consistent with the results from other
models \cite{scho1980, lavr1998,prasher2009}, while it is linear dependent on the interface coupling in 1D junctions. This
is due to the fact that in 3D low frequency region contributes
relatively little to the conductance as the density of states is low
there. If the interface coupling increases further, that is
$k_{12}/k_{12m}>0.1 $, out of the weak interface coupling limit, all
the modes contribute to the transmittance, the conductance is no
longer proportional to the square of the interface coupling, and the
slope continuously decreases. In some intermediate ranges the
conductance is approximately proportional to the interfacial
coupling (see Fig.~\ref{fig9}(b)), which is consistent with the
results from molecular simulation approach \cite{hu2009}. For
stronger coupling the conductances for the 1D case and 3D one have
similar behaviors, the slope of both cases will decrease
continuously to be zero at point $k_{12m}$, where the conductance
will be maximized and then decrease slightly to a limiting value.

\subsection{Thermal transport in extended junctions}

\begin{figure}[t]
\includegraphics[width=3.0 in,angle=0]{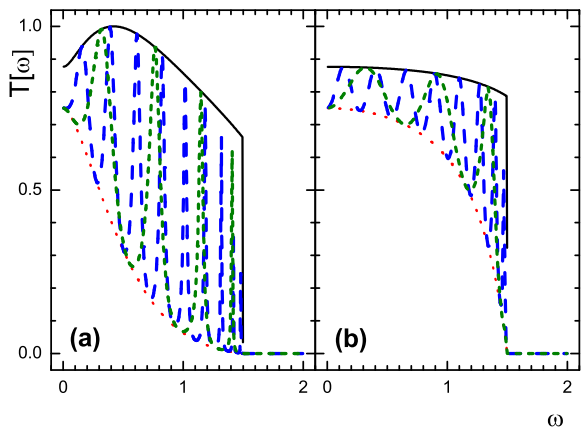}%
\caption{ \label{fig10} (color online)  The
transmission coefficient of the two-junction atomic chains.
Parameters: $k_1=1.0,\,m_1=1.0,\,
k_2=0.9,\,m_2=1.6,\,k_3=4.5,\,m_3=2.0$, The solid, dotted, dashed
and shot dashed lines correspond to maximum transmission, minimum
transmission, $N_c=4$ and $N_c=9$, respectively. The interface
couplings are different: (a) $k_{12}=0.3,k_{23}=0.7$; (b)
$k_{12}=1.0,k_{23}=4.5$. }
\end{figure}
\begin{figure}[t]
\includegraphics[width=3.0 in,angle=0]{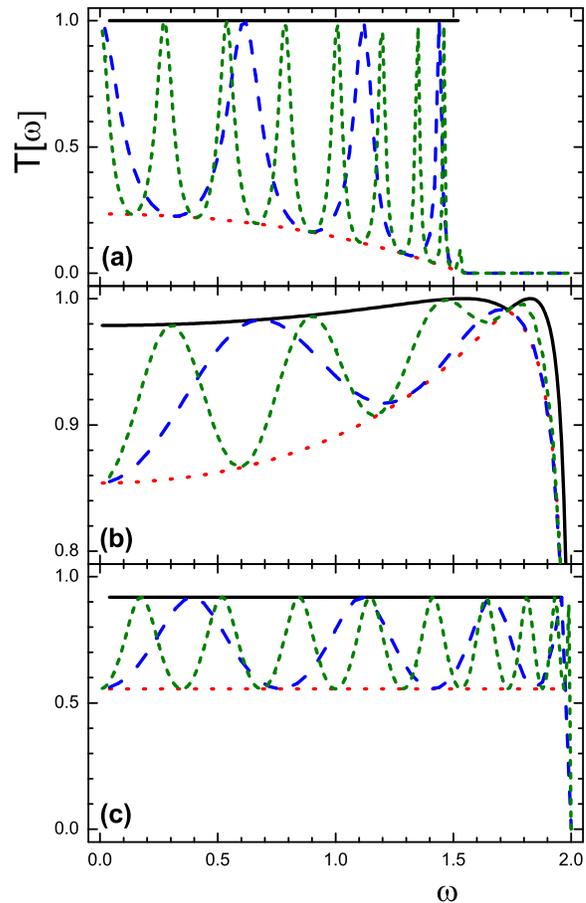}%
 \caption{\label{fig11} (color online) The
transmission coefficient of the two-junction atomic chains. Here,
$k_1=1.0,\,m_1=1.0$. The solid, dotted, dashed and shot dashed
lines correspond to maximum transmission, minimum transmission,
$N_c=4$ and $N_c=9$, respectively. (a)
$k_2=3.0,\,m_2=5.0,\,k_3=1.0,\,m_3=1.0,\,k_{12}=k_{23}=1.0$; (b)
$k_2=3.0,\,m_2=1.0,\,k_3=5.0,\,m_3=1.0,\,k_{12}=k_{12m}=1.5,\,k_{23}=k_{23m}=3.75$;
(c)
$k_2=3.0,\,m_2=3.0,\,k_3=5.0,\,m_3=5.0,\,k_{12}=k_{12m}=1.5,\,k_{23}=k_{23m}=3.75$.}
\end{figure}

Now we focus on a case where the junction is extended and involves a
center part. The overall behavior of the transmission is the
combination of the transmission behavior in single point-junction
case and the oscillatory behavior due to phonon interferences
arising form multiple scattering.   We show the transmission
coefficient as a function of frequency of an arbitrary case in
Fig.~\ref{fig10}(a). Here, the three chain parts have different
masses and spring constants, and the interface coupling is not
special. From the analytical expression of Eq.~(\ref{eqtran2}), we
plot curves of the maximum transmission and minimum transmission,
$N_c=4$ and $N_c=9$. The transmission oscillates between the envelop
lines of maximum and minimum transmission. The maximum transmission
line will increase first, and the minimum transmission line will
monotonically decrease with frequency. However for interface
coupling that is the same with the leads, the two envelop lines will
monotonically decrease, which can be seen in Fig.~\ref{fig10}(b).

\begin{figure}[t]
\includegraphics[width=3.0 in,angle=0]{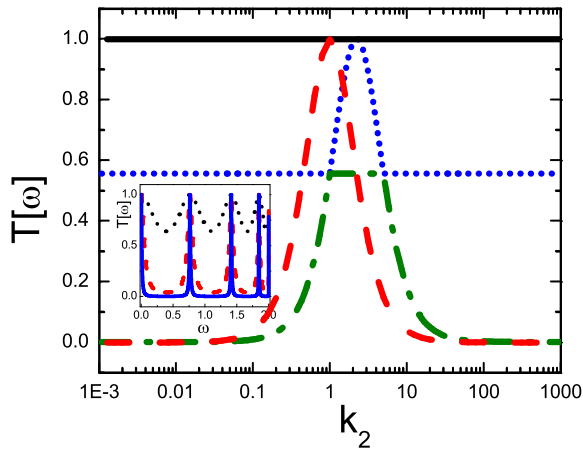}%
 \caption{\label{fig12} (color online) The maximum and
minimum transmission coefficient of the two-junction atomic chains.
Here, $k_1=1.0,\,m_1=1.0$. The solid, dashed lines correspond to
maximum transmission and minimum transmission
$k_3=1.0,\,m_3=1.0$, respectively; the dotted and dash-dotted lines
correspond to maximum transmission and minimum transmission
$k_3=5.0,\,m_3=5.0$, respectively.  The inset shows the transmission
coefficient with frequency for different $k_2$.
$k_1=k_3=1.0,\,m_1=m_3=1.0$. The dotted, dashed, and solid lines
correspond to $k_2=0.5$, 0.1, and 0.02 respectively. For all the
curves, $m_2=k_2$ and $k_{12}=k_{12m}, k_{23}=k_{23m}$. }
\end{figure}

For some special cases, the transmission coefficient in the
frequency domain has interesting phenomena, which are shown in
Fig.~\ref{fig11}. In Fig.~\ref{fig11}(a), the transmission for the
case of two identical leads is shown. In this case, the maximum
transmission is equal to one, the infinite-long wavelength phonon
and the  resonance mode can transmit fully through the center part.
The minimum transmission is very low, indicating efficient
destructive interference. Figure~\ref{fig11}(b) shows the
transmission when all three parts are different and connected by
interface couplings $k_{12m}$ and $k_{23m}$. We find that overall
trend for the maximum and minimum transmission lines is increasing
first, then decreasing.  If, in addition, the ratios of $k_i/m_i$
are the same for three parts, then the maximum and minimum
transmission are constants in the whole frequency range, and the
transmission coefficient through finite-size center part oscillate
between the two constants, which can be clearly seen in
Fig.~\ref{fig11}(c). Therefore, we can use the above properties of
transmission to design the frequency filters. Figure~\ref{fig12}
shows the maximum and minimum transmission coefficient for the
filter. If the spring constant of the center part is very different
from the ones of the the two leads, the oscillatory peak
is sharp, and transmission for most of the frequency will tend to
zero, only few resonant frequency can be transmitted.  This finding
provides guidelines for the design of selective frequency filters.

\section{Verification by Nonequilibrium Green's Function Method }
\label{secnegf} The NEGF method is an exact approach to study  the
ballistic thermal transport through junctions. Following the
discussion in Sec.~\ref{secmod}, if we use a transformation for the
coordinates, $u_j  = \sqrt {m_j } x_j $, which is called the
mass-normalized displacement, then the Hamiltonian can be written as
\begin{equation}
H  = \sum\limits_{\alpha=1,2,3} H_{\alpha}
+ \sum\limits_{\beta=1,3}
{U_\beta ^T V_{\beta, 2} U_2 },
\end{equation}
where $H_\alpha  = \frac{1}{2}\left(P_\alpha ^T P_\alpha + U_\alpha
^T K_\alpha  U_\alpha\right)$. $K_\alpha$ is the mass-normalized
spring constant matrix,  and $V_{12} = (V_{21})^T$ is the coupling
matrix of the left lead to the central region and  similarly for
$V_{23}$ is the coupling matrix of the right lead to the central
region. As stated in Ref.~\cite{wangjs2006}, the element of the coupling
matrix $V_{\alpha,\beta}^{ij}$ is equal to $-k_{ij}/sqrt{m_im_j}$
which corresponding to the coupling between the $i_{\rm th}$ atom
in region $\alpha$ and the $j_{\rm th}$ atom in region $\beta$.
\begin{figure}[t]
\includegraphics[width=3.0 in,angle=0]{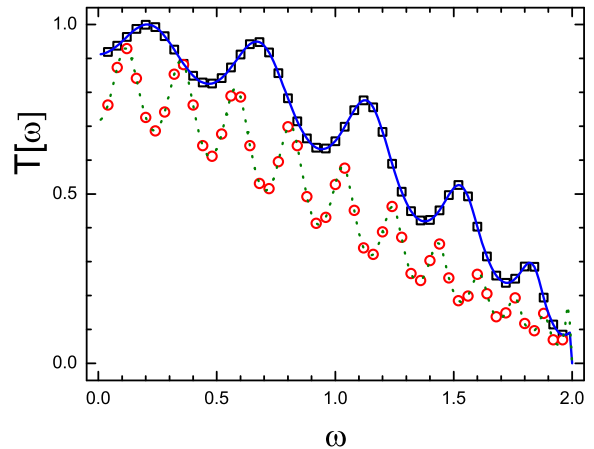}%
 \caption{ \label{fig13com}(color online)  The comparison of the
results from scattering boundary method and nonequilibrium Green's
function method for the transmission coefficient in two-junction
atomic chains. The square curve and solid line correspond the
parameters: $N_c=6,\, k_1=1.0, \,m_1=1.0,\,k_2=1.5,\,
m_2=1.3,\,k_3=2.0,\, m_3=1.7,\, k_{12}=1.3,\, k_{23}=0.8$; the
circle curve and dashed line correspond the parameters: $N_c=13,\,
k_1=1.0, \,m_1=1.0,\,k_2=1.5,\, m_2=1.3,\,k_3=4.0,\, m_3=2.7,\,
k_{12}=1.3,\, k_{23}=0.8$. The square and circle curves are the
results from nonequilibrium Green's function method; The solid and
dash lines are the results from scattering boundary method.}
\end{figure}

We can use the nonequilibrium Green's function method
\cite{wangjs2008} to study the thermal transport in the atomic
chain. We define the contour-ordered Green's function as
\begin{equation}
G^{\alpha \beta } (\tau ,\tau ') \equiv  - \frac{i}{\hbar}\left\langle
 {\mathcal{T}\, U_\alpha  (\tau )U_\beta  (\tau')^T } \right\rangle,
\end{equation}
where $\alpha$ and $\beta$ refer to the region that the
coordinates belong to and $\mathcal{T}$ is the contour-ordering operator.
Then the equations of motion of the  Green's function
can be derived.  In particular, the retarded Green's function for
the central region in frequency domain is
\begin{equation}
G^r [\omega ] = \Bigl[(\omega  + i\eta )^2  - K_2  - \Sigma^r [\omega ] \Bigr]^{-1}.
\end{equation}
Here, $\Sigma^r=\sum\limits_{\alpha=1,3} {\Sigma_\alpha ^r}$, and $
\Sigma_\alpha =V_{2, \alpha} g_\alpha V_{\alpha,2}$ is the
self-energy due to interaction with the heat bath, $
g_\alpha^r=[(\omega+i\eta)^2-K_\alpha]^{-1}$.   And in the advanced
Green's function $ G^a=(G^r)^\dag$, the transmission coefficient can
be calculated by the so-called Caroli formula as
\begin{equation}\label{eqcarl}
 T_{\beta \alpha } [\omega ] = {\rm{Tr}}(G^r \Gamma _\beta
G^a \Gamma _\alpha ),
\end{equation}
where $ \Gamma _\alpha = i\bigl(\Sigma _\alpha ^r
[\omega ] - \Sigma _\alpha ^a [\omega ]\bigr).$

For single-junction atomic chains, if we regard  the two atoms in
the interface (atom 0 and atom 1) as the center part, then we can
still use the formulae above to study the phonon transmission
leading to the exact formula yielding the same result with the one
obtained from the scattering boundary method. In Appendix A, We give
the analytical proof of this fact.

For two-junction atomic chains, according to the NEGF formulas, we do the numerical calculation and plot the curves of the transmission coefficient as a function of frequency and compare them to the results obtained the scattering boundary method (see Fig.~13). We find that for any arbitrary case, the results from the NEGF method and the scattering boundary method are exactly the same.
If there is no many-body interaction, that is, for the ballistic thermal transport the scattering  matrix approach and the Green's function method  give the same results.  These two methods are equivalent, which has been proved from other points of view in Refs.~\cite{khomyakov2005,harbola2008}.
\section{Conclusion}
\label{seccond}

In this paper, we study the ballistic interfacial thermal transport in atomic junctions, we give the analytical simple formulae Eq.~(\ref{eqtran1}), Eq.~(\ref{eqrflt}) and Eq.~(\ref{eqtran2}) for the transmission of one-junction and two-junction cases, which are consistent with the results from the NEGF method.

For one-junction case, we find the transmission and conductance are maximized when the interface spring constant equals to the harmonic average of the two spring constants of the leads. At the point near $k_{12}=k_{12m}$, the transmission $T[\omega]$ is a constant if $k_2/m_2=k_1/m_1$; if not equal, in the range between $k_1/m_1=k_2/m_2$ and $k_1 m_1=k_2 m_2$, the transmission coefficient increases first then decreases with the increasing of frequency, otherwise the transmission monotonically decreases as the frequency increasing.
For weak interface coupling, the cutoff frequency and the interface conductance for 1D chain is linear dependent with the interface coupling strength.

Because of different density of states, we change the formula of conductance to mimic the thermal transport in 3D junctions. In weak interface coupling limit, we find that the conductance is proportional to the square of the interface coupling, which is consistent with the results from other models. The slope of the conductance as function of interfacial coupling strength decreases continuously from two to zero, in certain range of which, the conductance is linear proportional to the interface coupling, which are consistent with the results of other molecular simulations.

For two-junction case, the transmission will oscillate with frequency in the envelop lines of maximum and minimum transmission which are determined by the one-junction picture.  The transmission sometimes oscillates between two decreasing envelop lines, sometimes between two increasing envelop curves, or between two constants, etc.

\section*{Acknowledgements}
  P. K. is supported by the U.S. Air Force Office of Scientific Research Grant No. MURI FA9550-08-1-0407. J.-S. W. acknowledge
support from a NUS research grant R-144-000-257-112.

\appendix
\section{Analytical proof of the equality of the two methods for one junction  }

In this appendix we give the analytical proof for the equality of the scattering boundary method and the non-equilibrium Green's function approach for the one-junction atomic chains.

From the scattering boundary method, we obtain the transmission Eq.~(\ref{eqtran1t}) and Eq.~(\ref{eqtrsm}), that is

\begin{widetext}
\begin{equation}
T[\omega ] = \frac{{\sqrt {4k_2 m_2  - \omega ^2 m_2^2 } }} {{\sqrt {4k_1 m_1  - \omega ^2 m_1^2 } }}\big|\frac{{ - k_{12} k_1 (\lambda _1  - 1/\lambda _1 )}}
{{(k_1  - k_{12}  - k_1 /\lambda _1)(k_2  - k_{12}  - k_2 /\lambda _2 ) - k_{12}^2 }}\big|^2,
\end{equation}
\end{widetext}

From the dispersion relation Eq.~(\ref{eqdspr}), we can obtain
\begin{equation}
k_j-k_j/\lambda_j=\omega ^2 m_j-k_j(1-\lambda_j)
\end{equation};
and
\begin{equation}
k_j^2|\lambda_j-1/\lambda_j|^2=\omega ^2 (4 k_j m_j-\omega ^2 m_j^2)
\end{equation},
So we can get
\begin{widetext}
\begin{equation}\label{eqtransbm}
T[\omega ] = \frac{k_{12}^2 \omega ^2 \sqrt {4k_1 m_1  - \omega ^2 m_1^2 } \sqrt {4k_2 m_2  - \omega ^2 m_2^2 } }{\big|
[\omega ^2 m_1-k_1(1-\lambda _1 )-k_{12}][\omega ^2 m_2-k_2(1-\lambda _2 )-k_{12}] - k_{12}^2 \big|^2}.
\end{equation}
\end{widetext}
Using the NEGF formulae, we regard the two
atoms in the interface (atom 0 and atom 1) as the center
part 0, then the dynamic matrix of the center as
\begin{equation}
  K_0=  \left(
{\begin{array}{*{20}c} \frac{k_1+k_{12}}{m_1}& \frac{-k_{12}}{\sqrt{m_1 m_2}} \\ \frac{-k_{12}}{\sqrt{m_1 m_2}}  &  \frac{k_{12}+k_{2}}{m_1} \\
\end{array}} \right).
\end{equation}
And the coupling matrices between the leads (parts 1 and 2) and the center (part 0) are $V_{01}=(k_1/m_1\,,\,0 )^T$ and $V_{02}=(0\,,\,k_2/m_2)^T$, and according to Ref.~\cite{wang2007}, we can obtain the surface Green's function as
\begin{equation}
    g_i^r=-\frac{m_i \lambda_i}{k_i},
\end{equation}
here, $i=1,2$ corresponds to the left and right lead.
Then we can get the self energy ($\Sigma^r=V_{01} g_1^r V_{10}+V_{02} g_2^r V_{20}$) as
\begin{equation}
\Sigma^r= \left(
{\begin{array}{*{20}c} -\frac{k_1 \lambda_1}{m_1}& 0 \\ 0 &  -\frac{k_2 \lambda_2}{m_2}\\
\end{array}} \right).
\end{equation}
Thus we can calculate the retarded Green's function of the center $G^r=(\omega^2 I-K_0-\Sigma^r)^{-1}$, which reads as
\begin{equation}
G^r=\left(
{\begin{array}{cc} A_1& B \\ B &  A_2\\
\end{array}} \right)^{-1}=\frac{1}{\Delta}\left(
{\begin{array}{cc} A_2& -B \\ -B &  A_1\\
\end{array}} \right),
\end{equation}
here, $I$ is two-dimensional identity matrix and
\begin{eqnarray}
 A_i=\omega^2-\frac{k_i}{m_i}(1-\lambda_i)-\frac{k_{12}}{m_i}; \\
 B=\frac{k_{12}}{\sqrt{m_1 m_2}};\; \Delta=A_1 A_2 -B^2.
\end{eqnarray}
The advanced Green's function $G^a$ equals to $(G^r)^\dag$.
And from the self energy we can get
\begin{equation}
\Gamma_1=\left(
{\begin{array}{*{20}c} C_1& 0 \\ 0 & 0\\
\end{array}} \right); \; \Gamma_2=\left(
{\begin{array}{*{20}c} 0& 0 \\ 0 &  C_2\\
\end{array}} \right),
\end{equation}
here, $C_i=\frac{\omega}{m_i} \sqrt {4k_i m_i  - \omega ^2 m_i^2 }$.
Therefore, we can calculate the transmission coefficient from the Caroli formula Eq.~(\ref{eqcarl}), at last we obtain
\begin{equation}
 T[\omega]=Tr(G^r \Gamma_1 G^a \Gamma_2)=\frac{B^2 C_1 C_2}{\Delta \Delta^*}=\frac{B^2 C_1 C_2}{|A_1 A_2- B^2|^2}
\end{equation}
Inserting the values of $A_i, B$ and $C_i$,  we get exactly the same
result with Eq.~(\ref{eqtransbm}). Therefore, the results from the
scattering boundary method and non-equilibrium Green's function
approach are equivalent.

\end{document}